\documentclass[sigconf,review=false,anonymous=false]{acmart}
\usepackage{array}
\usepackage{graphicx}
\usepackage{clrscode}
\usepackage{balance}
\usepackage{subfigure}
\usepackage{multirow}
\usepackage{booktabs}
\usepackage{adjustbox}
\usepackage[justification=justified,skip=5pt]{caption}
\usepackage{float}
\usepackage{color}
\usepackage{soul}
\usepackage{mdframed}
\usepackage{amsopn}
\usepackage{mathrsfs}
\usepackage{mathtools}
\usepackage{amsmath}
\usepackage{arydshln}
\usepackage{hyperref}
\usepackage{multicol}
\usepackage{blkarray}
\usepackage{enumerate}
\usepackage{courier}
\usepackage{rotating}
\usepackage{booktabs}
\usepackage{diagbox}
\usepackage{fancybox}
\usepackage{minibox}
\usepackage{cases}
\usepackage{bm}
\usepackage{enumitem}
\usepackage[lined, ruled, commentsnumbered, linesnumbered]{algorithm2e} %
\usepackage{enumitem}
\setlist[itemize]{leftmargin=*}

\AtBeginDocument{%
  \providecommand\BibTeX{{%
    \normalfont B\kern-0.5em{\scshape i\kern-0.25em b}\kern-0.8em\TeX}}}

\acmConference{SIGIR '23}{July 23rd - 27th, 2023}{Taipei}

\begin{document}

\title{Automated Data Denoising for Recommendation}


\author{Yingqiang Ge}
\affiliation{%
  \institution{Rutgers University}
  \country{}
}
\email{yingqiang.ge@rutgers.edu}
\thanks{Work was done when Yingqiang worked as a summer intern at Amazon in 2022.}

\author{Mostafa Rahmani}
\affiliation{%
  \institution{Amazon.com, Inc.}
  \country{}
}
\email{mostrahm@amazon.com}

\author{Athirai Irissappane}
\affiliation{%
  \institution{Amazon.com, Inc.}
  \country{}
}
\email{athirai@amazon.com}

\author{Jose Sepulveda}
\affiliation{%
  \institution{Amazon.com, Inc.}
  \country{}
}
\email{joseveda@amazon.com}

\author{James Caverlee}
\affiliation{%
  \institution{Texas A \& M University}
  \country{}
}
\email{caverlee@gmail.com}

\author{Fei Wang}
\affiliation{%
  \institution{Amazon.com, Inc.}
  \country{}
}
\email{feiww@amazon.com}


\begin{abstract}

In real-world scenarios, most platforms collect both large-scale, naturally noisy implicit feedback and small-scale yet highly relevant explicit feedback. 
Due to the issue of data sparsity, implicit feedback is often the default choice for training recommender systems (RS), however, such data could be very noisy due to the randomness and diversity of user behaviors. 
For instance, a large portion of clicks may not reflect true user preferences and many purchases may result in negative reviews or returns.
Fortunately, by utilizing the strengths of both types of feedback to compensate for the weaknesses of the other, we can mitigate the above issue at almost no cost.
In this work, we propose an Automated Data Denoising framework, \textbf{\textit{AutoDenoise}}, for recommendation, which uses a small number of explicit data as validation set to guide the recommender training.
Inspired by the generalized definition of curriculum learning (CL), AutoDenoise learns to automatically and dynamically assign the most appropriate (discrete or continuous) weights to each implicit data sample along the training process under the guidance of the validation performance. 
Specifically, we use a delicately designed controller network to generate the weights, combine the weights with the loss of each input data to train the recommender system, and optimize the controller with reinforcement learning to maximize the expected accuracy of the trained RS on the noise-free validation set. 
Thorough experiments indicate that AutoDenoise is able to boost the performance of the state-of-the-art recommendation algorithms on several public benchmark datasets.

\end{abstract}





\maketitle

\section{Introduction}
Recommender systems (RS) are an essential part of modern life, which are widely deployed in almost every corner of our daily routines 
and facilitate the human decision-making process by providing relevant suggestions. 
Various techniques have been adapted to enhance the capabilities of deep recommendation systems with the common goal of predicting more accurate user preferences, including but not limited to meticulously 
selecting user/item interaction features \cite{khawar2020autofeature, liu2020autofis}, or proposing novel loss functions \cite{zhao2021autoloss,10.1145/3477495.3531941}.
However, very few works doubt the reliability of the data source.

General recommender systems focus on modeling the compatibility between users and items, based on historical user-item interaction (e.g. clicks, purchases, likes).
Existing literature usually separate user-item interaction into two categories \cite{he2017neural,ge2022survey,hu2008collaborative,10.1145/3209978.3210007}: explicit feedback and implicit feedback. 
On one hand, explicit feedback, such as ratings and reviews, are collected when users actively and explicitly tell the system about their preferences for an item, which are usually small-scale yet highly relevant.
On the other hand, implicit feedback, such as user clicks or purchases, are passively recorded when users interact with the website interface, so they are easily collectable and large-scale but can only implicitly reflect the part of the users' preferences, as they are easily affected by the first impression of users and other factors, such as click baits \cite{wang2021clicks}, position bias \cite{10.1145/3331184.3331269}, etc.
In order to mitigate the data sparsity issue during model training, most deep recommenders utilize implicit data as input, however, prior work \cite{hu2008collaborative,10.1145/3209978.3210007,10.1145/3298689.3347037} points out that it is more challenging to utilize implicit data since there is an inevitable gap between implicit feedback and the actual user satisfaction due to the prevailing presence of noisy interactions (a.k.a, false-positive interactions), where the users may dislike certain interacted item.
For example, in e-commerce platforms, large portions of clicks do not align well with user preferences, and many purchases end up with negative reviews or being returned.
Moreover, existing studies \cite{10.1145/3298689.3347037,wang2021denoising} have further demonstrated the detrimental effect of such noisy data samples through both offline evaluation and online tests.
However, even though it is of critical significance to account for the inevitable noisy nature of implicit feedback for recommender training, little work has been done yet.

Existing efforts dedicated to denoising user-item interactions can be separated into two categories: 1) leveraging additional feedback (e.g., dwell time, gaze pattern, favorite, skip) either to predict then remove the noisy data or to incorporate these feedback into the training process \cite{liu2010understanding,yang2012exploiting,10.1145/3298689.3347037}; 2) utilizing domain expert knowledge to eliminate the effects of false-positive interactions, such as setting threshold for losses \cite{wang2021denoising}. 
These represent two extreme solutions: the former always use explicit data, which needs additional feedback and extensive manual label work, making it infeasible in most cases; the latter does not use any explicit data since it is based on strong assumptions, while such assumptions are made by domain experts for specific datasets and recommenders.
Moreover, both of them neglect the fact that the platforms usually have both large-scale but noisy implicit feedback, and small-scale yet highly relevant explicit feedback.
Explicit data suffers from the scale issue while implicit data may not, and implicit data is noisy but explicit data may not.
Thus, leveraging each other’s strengths to make up for the other’s weaknesses provides a natural, feasible and fully automatic solution for the implicit data denoising problem.


To this end, we propose an Automated Data Denoising framework, named \textit{AutoDenoise}, for recommenders, which exploits a small volume of noise-free data (i.e., those true-positive user-item interactions) to guide the recommenders to recommend more satisfying (or true-positive) items to users. 
Inspired by generalized curriculum learning (CL) \cite{bengio2009curriculum, wang2021survey}, AutoDenoise learns to automatically and dynamically select the most suitable examples, or assign the most appropriate weights to each sample along the training process under the guidance of the performance on the noise-free data, where the former is named as \textit{Hard AutoDenoise} (AutoDenoise-H) and the latter is called \textit{Soft AutoDenoise} (AutoDenoise-S). 
Moreover, unlike traditional CL, our framework allows examples to appear many times (i.e., receive weights greater than one) so as to make a clearer difference between different samples, which helps to generalize data weighting, filtering, and fine-tuning schemes.
Technically, following neural architecture search (NAS) in \cite{pham2018efficient,zoph2016neural}, we use a reinforcement learning approach involving a learned agent, named controller, whose task is to learn to assign weights to each implicit data sample.
Under the guidance of noise-free data samples, the domain knowledge of true-positive interactions is automatically learned by the controller network.
Specifically, we use a meticulously designed controller network to generate (hard or soft) weights, combine the weights with the loss of each input data into a weighted loss to train the recommender system, and optimize the controller with REINFORCE \cite{williams1992simple} to maximize the expected accuracy of the trained RS on the clean and noise-free data.
In our implementation, to make more use of these noise-free data, we also turn it into a validation set, so its recommendation accuracy can be further used for both hyper-parameter tuning and controller's reward signal.





The contributions of this work can be summarized as follows:
\begin{itemize}
    \item We propose an end-to-end framework, AutoDenoise, which is able to automatically and dynamically select the most appropriate data instances (or assign the most
appropriate weights to each sample) for RS training to help reduce the noisy signals and improve the recommendation performance;
    
    \item We utilize the expected accuracy on the small-scale noise-free validation set to guide the training process, leveraging the strengths of implicit data and explicit data to compensate for each other’s weaknesses, 
    which is a natural and elegant solution;
    
    \item Extensive experiments on three benchmarks using both classical and deep recommendation models validate the effectiveness and generalizability of AutoDenoise.
\end{itemize}

\section{Related Work}\label{sec:related_work}
In this section, we will briefly introduce some background knowledge about noise in recommendation, curriculum learning in recommendation and automated machine learning in recommendation.

\subsection{Noise in Recommendation}

Generally, there are two types of noises in recommender systems \cite{10.1145/1111449.1111477}: malicious noise (shilling noise, e.g., injection attacks) and non-malicious noises (natural noise, e.g., human errors). 
The former is usually the result of malicious users deliberately manipulating the predicted scores of certain items, while the latter is related to random user behaviors during the selection of items. 
In collaborative filtering recommendation systems, there have already been many approaches focusing on detecting malicious noise,
while, given that natural noise is often hidden in the user’s behavior, there are only a few studies related to this problem. 
For instance, \citeauthor{li2013noisy} \cite{li2013noisy} utilized a novel real-time quadratic optimization algorithm for identifying and removing noisy non-malicious users (NNMUS), which deals with natural noise through detecting noise but not malicious users if the user’s rating for closely related items has the same score. 
Moreover, some existing work collected the various users’ feedback (e.g., dwell time \cite{10.1145/2556195.2556220}, gaze patterns \cite{zhao2016gaze}, and skip \cite{fox2005evaluating}) and the item characteristics \cite{10.1145/3209978.3210007} to predict the user’s satisfaction,
while other approaches \cite{liu2010understanding,yang2012exploiting,10.1145/3298689.3347037} directly incorporated additional feedback into training.
For example, \citeauthor{10.1145/3298689.3347037} \cite{10.1145/3298689.3347037} used three types of items: ``click-complete'', ``click-skip'', and ``non-click'', to train the recommendation model, where the last two types of items are both treated as negative samples but with different weights.
All these methods need additional feedback and extensive manual label work, e.g., users have to tell their satisfaction of each interaction. Furthermore, the quantification of item quality largely relies on the manual feature design and the labeling of domain experts \cite{10.1145/3209978.3210007}. 
The unaffordable labor cost impedes the practical application of these methods, especially in the scenarios where item pools change over time, such as news recommendation, movie recommendation, etc.

On the contrary, our proposed framework, AutoDenoise, takes advantage of the fact that the platforms usually have both large-scale noisy implicit feedback and small-scale yet noise-free explicit feedback, which makes it possible to leverage each other's strengths to make up for the other’s weakness in the data denoising problem, leading to a feasible and elegant solution.

\subsection{Curriculum Learning in Recommendation}

The idea of curriculum learning (CL) was popularized by \citet{bengio2009curriculum}, who viewed it as a way to improve convergence by presenting heuristically identified easy examples first \cite{bengio2009curriculum}.
The core idea of CL is that easier instances should be involved first in the model learning, and then more complex ones are gradually considered. 
This tactic is empirically evaluated to be beneficial for the learning process to alleviate the bad local minimum, leading to the superior performance in some tasks \cite{SARAFIANOS201894}.
To make a recommender model understand user preferences more precisely, various studies have suggested curriculum learning during model training \cite{ma2018point, hu2008collaborative, zhang2018discrete}. 
For example, in movie recommendation, a self-paced learning (SPL) based reweighting \cite{kumar2010self} strategy was proposed by \citet{zhang2018discrete} to select the samples for reweighting under a predefined weighting scheme.
As another example, \citet{liu2021self} designed a bounded SPL learning paradigm with a parameter to control how many instances will be finally induced in the model learning, which tries to learn the model mainly on clean data and exclude noisy instances.

Moreover, among works on handling noisy data in recommendation, the closest to ours is \citet{wang2021denoising}, which proposed two strategies to deal with noisy data: 1) truncating the loss values of hard interactions to 0 with a dynamic threshold; 2) assigning hard interactions with smaller weights based on a variant of Focal Loss \cite{lin2017focal}. Nevertheless, this setting has two major drawbacks: first, it still uses fixed weighting functions designed with domain expert knowledge, which is time consuming and inflexible; second, it exploits a strong assumption that deep models will learn the easy and clean patterns in the early stage, which has been proven to be not always true in many other cases \cite{48346,zhou2018minimax}. Our proposed framework is neither based on the ``easier first'' assumption nor the ``harder first'' assumption \cite{wang2021survey}, instead it learns to allocate the most appropriate training data in a trial and error fashion, i.e., reinforcement learning. We also replace the fixed predefined weighting function with a learnable deep neural network, which enhances its ability of generalization.

\subsection{AutoML in Recommendation}
The proposed framework is closely related to the field of Automated Machine Learning (AutoML) as it focuses on automating data processing tasks such as data selection and reweighting. In the light of this, we provide a brief overview of how AutoML is applied to the field of recommendation.
 
AutoML techniques have been widely introduced in recommendation systems. Specifically, existing works mainly focused on three research directions: 1) the design of the embedding layer, 2) the selection of feature interaction patterns, and 3) the generation of loss function.
For the embedding layer design, several works \cite{joglekar2020neural,liu2021learnable,zhao2020memory} proposed novel methods to automatically select the most appropriate embedding size for different feature fields. For example, \citet{liu2021learnable} proposed to dynamically search embedding sizes for users and items based on their popularity by introducing a novel embedding size adjustment policy network (ESAPN). 
Similarly, \citet{ginart2021mixed} proposed to use mixed dimension (MD) embedding layer, which consists of variable embedding sizes for each feature,
and \citet{cheng2020differentiable} proposed to perform embedding dimension selection with a soft selection layer, making the dimension selection more flexible. 
For feature selection, \citeauthor{luo2019autocross} proposed AutoCross in \cite{luo2019autocross}, which enables explicit high order feature interaction search on a tree-structured search space by implementing greedy beam search. 
In \cite{tsang2020feature}, the authors proposed to interpret feature interactions from a source recommender model and then encode these interactions in a target recommender model, where both source and target models are black-box models.
For loss function generation, there have been only a few pioneering works, such as \cite{zhao2021autoloss,10.1145/3477495.3531941}. For instance, \citet{10.1145/3477495.3531941} proposed an automatic loss function generation framework, AutoLossGen, which is able to generate loss functions directly constructed from basic mathematical operators without prior knowledge on the loss structure.

\section{Problem Formulation}\label{sec:problem}
In this section, we first provide an overview of the general recommendation problem, and then propose to integrate curriculum learning into the recommendation task to mitigate the impact of false-positive training samples.

\subsection{Recommendation Formulation}
Suppose we have a user set with $m$ users denoted as $\mathcal{U}$, an item set $\mathcal{V}$ with $n$ items and their interaction set $\mathcal{D}=\{(u,v,y_{uv},t)|u \in \mathcal{U}, v \in \mathcal{V}, y_{uv} \in \{0,1\}, t \in \textit{R}^+\}$,  $y_{uv}$ represents whether the user $u$ interacted with item $v$, and $t$ is the timestamp.
All users share an embedding matrix $\textbf{U} \in \mathbb{R}^{ m \times d}$, and items share an embedding matrix $\textbf{V} \in \mathbb{R}^{n \times d }$, where $d$ is the size of embedding vector. Accordingly, each user's or item's latent vector $\textbf{e}_{u} \in \mathbb{R}^{d}$ or $\textbf{e}_{v} \in \mathbb{R}^{d}$ 
is the corresponding row in the embedding matrix:
$$
\textbf{e}_{u}= \textbf{U}_{u} ; \quad \textbf{e}_{v}=\textbf{V}_{v} \text {. }
$$
The embedding matrices $\textbf{U}$ and $\textbf{V}$ for users and items are exactly what to learn in the training process, with supervised ground-truth of user-item interactions.
We define a model $f$, which is a function parameterized with $\theta$, to predict the user-item ranking score $\hat{y}_{u,v}$ for user $u$ and item $v$, as
\begin{equation}
    \hat{y}_{u,v} = f(\textbf{e}_{u}, \textbf{e}_{v}\mid Z, \Theta)
\end{equation}
where $\Theta = \{\textbf{U},\textbf{V},\theta\}$ contains all learnable model parameters, and $Z$ represents all other auxiliary information. 
Depending on the application, $Z$ could be rating scores, clicks, text, images, etc., and is optional in the recommendation model $f$ (we will omit $Z$ in the rest of our paper for simplicity).
The goal of a recommendation system can be stated as follows: learning a recommendation model from $\mathcal{D}$ so that it can capture user preferences and make high-quality recommendations.
\begin{equation}\label{eq:classic_loss}
\begin{aligned}
\mathcal{L}(\mathcal{D} | \mathbf{\Theta}) &=\frac{1}{|\mathcal{D}|} \sum_{x \in \mathcal{D}} \ell\big( f (\textbf{e}_{u}, \textbf{e}_{v} \mid \Theta ), y_{uv} \big)
\end{aligned}
\end{equation}
where $\ell(\cdot)$ can be any suitable loss functions, e.g., Binary Cross Entropy loss (BCE), Bayesian Personalized Ranking loss (BPR) \cite{bpr}, Softmax loss, etc. After training is complete, the top-$K$ items generated by the ranking score function $f$ except the interacted items are recommended to the user $u$. 


\subsection{Curriculum Learning for Recommendation}
Usually, Eq. \ref{eq:classic_loss} can be directly optimized by stochastic gradient descent (SGD) method. However, the noisy training samples will be treated equally as good samples for the update of the parameters during the process of model training, and the noisy training instances will greatly harm the effectiveness of the model. To overcome the above issue, curriculum learning (CL) mechanism is introduced. 


In this work, we mainly focus on automatic curriculum learning since traditional predefined curriculum learning methods heavily rely on human prior knowledge \cite{wang2021survey}.
Concretely, we aim to learn a reweighting of each input sample and minimize a weighted loss,
\begin{equation}\label{eq: spl_loss}
\begin{aligned}
\mathcal{L}^{\prime}(\mathcal{D} | \Theta, \textbf{w})
&= \frac{1}{|\mathcal{D}|} \sum_{(u,v) \in \mathcal{D}} w_{uv} \cdot \ell \big(f( \textbf{e}_{u}, \textbf{e}_{v} \mid \Theta ), y_{uv} \big) \\
\end{aligned}
\end{equation}

In classical CL, $\textbf{w} = [w_{uv}]^{|\mathcal{D}|} \in [0,1]^{|\mathcal{D}|}$ represents the weight vector, and each element $w_{uv}$ measures the easiness of each user-item pair $(u,v) \in \mathcal{D}$ to determine whether an instance is selected or not. 
Considering the large number of $w_{uv}$, we replace $\textbf{w}$ with a controller network $g$ in Eq. \ref{eq: spl_loss}, which is a deep neural network parameterized with $\Phi$. The new objective function is shown as follows,
\begin{equation}\label{eq: weighted_loss}
\begin{aligned}
\mathcal{L}^{\prime} (\mathcal{D} | \Theta, \Phi)
&= \frac{1}{|\mathcal{D}|} \sum_{(u,v) \in \mathcal{D}} g ( \textbf{e}_{u}, \textbf{e}_{v} \mid \Phi) \ell \big(f ( \textbf{e}_{u}, \textbf{e}_{v} \mid \Theta ), y_{uv} \big)
\end{aligned}
\end{equation}

Finally, our new optimization problem is,
\begin{equation}\label{eq: weighted_obj}
\begin{aligned}
\min_{\Theta, \Phi} \ & \mathcal{L}^{\prime} (\mathcal{D} | \Theta, \Phi) + \lambda_1 ||\Theta||^2 + \lambda_2 ||\Phi||^2
\end{aligned}
\end{equation}
where $\lambda_1 \in [0,1]$ and $\lambda_2 \in [0,1]$ are hyper-parameters used to control the weights between the three terms, and $||\Theta||^2$ and $||\Phi||^2$ represent L2 penalty. 
Most of the important symbols used in the paper can be referred in Table \ref{Table:notation}.

Moreover, without losing generality, we only formulate the problem based on collaborative filtering methods and focus on reweighting the training loss. 
However, the proposed framework can be applied to any differentiable recommendation models. For example, the user embedding $\textbf{e}_{u}$ can also be replaced with other hidden representations, such as the representation of user behavior sequences in sequential recommendation. We also demonstrate the generalizability of our framework by applying it to various recommendation algorithms in the experimental section.

\begin{table}[]
    \centering
    \begin{tabular}{c l}
    \toprule
      {\bfseries Symbol} & {\bfseries Description}\\
      \midrule
      $\mathcal{U}$ & The set of users in a recommender system\\
      $\mathcal{V}$ & The set of items in a recommender system\\
      $\mathcal{D}$ & The set of user-item interactions \\
      $m$ & The number of users\\
      $n$ & The number of items\\
      $u$ & A user ID in a recommender system\\
      $v$ & An item ID in a recommender system\\
      $\boldsymbol{U}$ & A user embedding matrix\\
      $\boldsymbol{V}$ & A item embedding matrix \\
      $\textbf{e}_{u}$ & User $u$'s embedding\\
      $\textbf{e}_{v}$ & Item $v$'s embedding\\
      $d$ & The dimension of user/item embedding\\
      $y_{uv}$ & Ground-truth value of the pair $(u, v)$\\
      $\hat{y}_{uv}$ & Predicted value of the pair $(u, v)$\\
      $K$ & The length of the recommendation list\\
      $f_\Theta(\cdot)$ & The recommendation model parameterized with $\Theta$\\
      $g_\Phi(\cdot)$ & The controller network parameterized with $\Phi$\\
      $w_{uv}$ & The weight assigned to pair $(u,v)$\\
      $\textbf{w}$ & The weight vector for all user-item interactions\\
      $\mathcal{R}_{val}$ & Reward on validation set\\
      \bottomrule
    \end{tabular}
    \vspace{5pt}
    \caption{Summary of the notations in this work.}
    \label{Table:notation}
\end{table}

\section{AutoDenoise}
With the above definitions, the objective of our AutoDenoise framework is to automatically and dynamically assign the most appropriate weights (discrete or continuous values) to each implicit data sample through learning $\Theta^*$ and $\Phi^*$ in Eq. \ref{eq: weighted_obj}. 
We will introduce in details on how to leverage reinforcement learning (RL) to optimize the controller network $g_{\Phi}(\cdot)$ and the recommendation model $f_{\Theta}(\cdot)$.

\subsection{Overall Procedure}
In AutoDenoise, the parameters to be optimized are from two networks, namely $f_\Theta$ and $g_\Phi$. 
For RS model, we perform stochastic gradient descent (SGD) to update $\Theta$ on the implicit training data, while for the controller, inspired by the success of neural architecture search (NAS) in \cite{pham2018efficient,zoph2016neural}, we apply the reinforcement learning algorithm to update $\Phi$, and the accuracy on the explicit validation dataset is used as the reward signal for policy gradient.
This forms a bi-level optimization problem \cite{anandalingam1992hierarchical}, where controller parameters $\Phi$ and RS parameters $\Theta$ are considered as the upper- and lower-level variables.
The optimization problem is formulated as:
\begin{equation}
\begin{split}
    \max\limits_{\Phi} \text{ } & \mathbb{E}_{ \textbf{w} \sim g({\mathcal{D}|\Phi, \Theta^*)}}[\mathcal{R}_{val}] \\
    \text{s.t.~~~} & \Theta^* = \text{arg}\min\limits_{\Theta}\mathcal{L}^{\prime}_{train}(\mathcal{D} | \Theta, \Phi^*)
\end{split}
\label{Eq:optimization}
\end{equation}
where $\mathcal{R}_{val}$ represents the reward (i.e., accuracy) on validation set.
To solve this problem, one can update $\Theta$ and $\Phi$ in an alternative manner. 
Specifically, $\Theta$ and $\Phi$ are alternately updated on training and validation data by minimizing the training loss $\mathcal{L}^{\prime}_{train}$ and maximizing the expected validation performance $\mathbb{E}_{ \textbf{w} \sim g({\mathcal{D}|\Phi, \Theta^*)}}[\mathcal{R}_{val}]$, respectively.

\begin{figure}[t]
\centering
\mbox{
\hspace{-25pt}
\centering
    \includegraphics[scale=0.3]{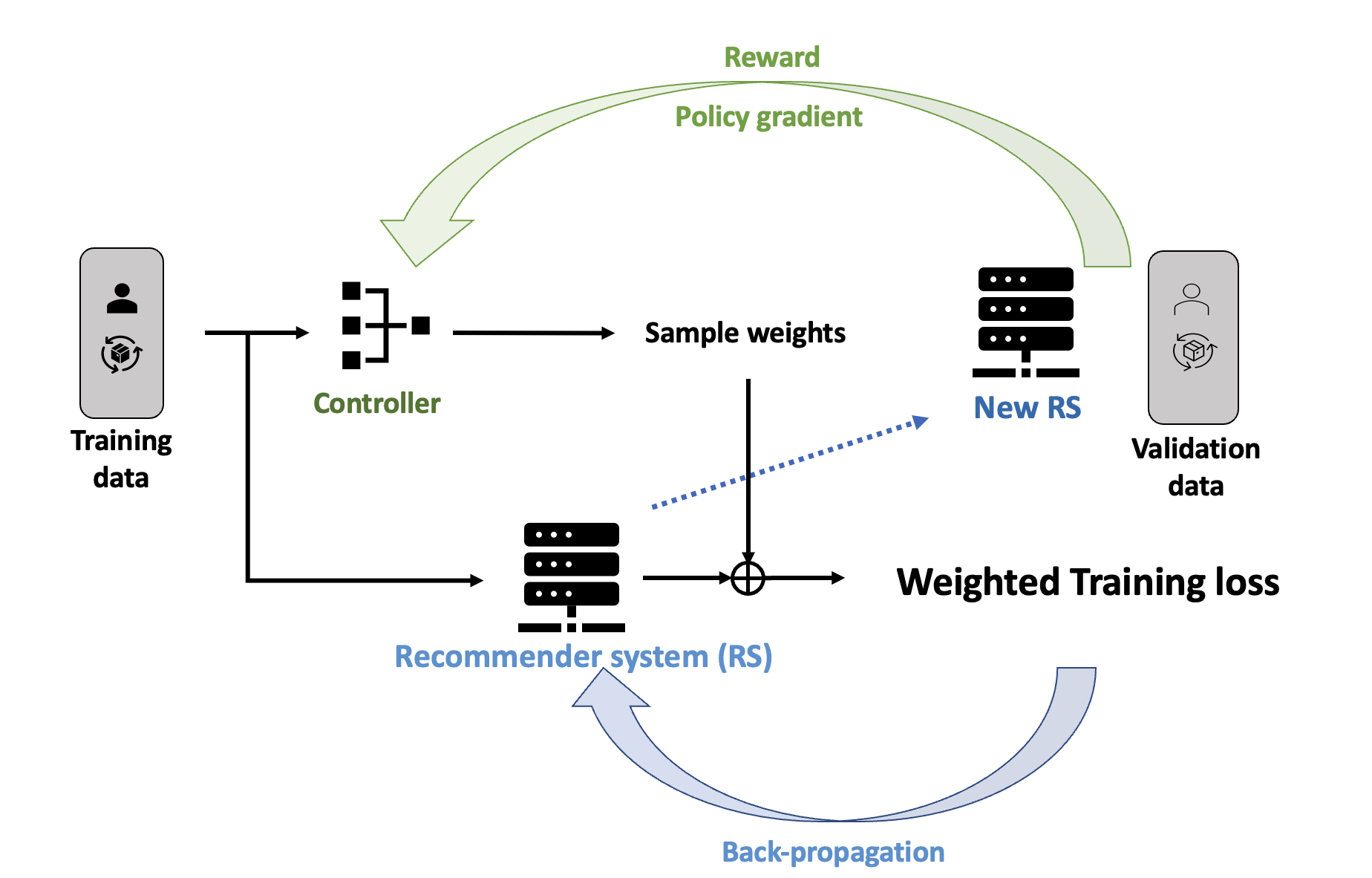}
}
\caption{Illustration of the proposed method.}
\label{fig:model}
\end{figure}

As shown in Fig. \ref{fig:model}, at each training epoch, the controller will dynamically select examples for training according to the recommender's feedback, and the trained recommender will result in an accuracy on the validation set. Using this accuracy as the reward signal, we can compute the policy gradient to update the controller. Moreover, the weight assignment is taken as the action in the RL schemes, and the feedback is taken as the state and reward.

\subsection{Training with REINFORCE}
In our setup, the environment is the recommender system and its training data, as illustrated in Fig. \ref{fig:model}.
The weight vector $\textbf{w}$ that the controller predicts can be viewed as an action for the training of the recommender.  We can use the accuracy on the validation set as the reward signal $\mathcal{R}_{val}$ and use reinforcement learning to train the controller. 
More concretely, to find the optimal architecture, we require our controller to maximize its expected reward, represented by $J(\Phi)$:
\begin{equation}
    J(\Phi) = \mathbb{E}_{\textbf{w} \sim g({\mathcal{D}|\Phi)}}[\mathcal{R}_{val}]
\end{equation}

Since the reward signal $\mathcal{R}_{val}$ is non-differentiable, we need to use a policy gradient method to iteratively update $\Phi$. In this work, we use the REINFORCE in \cite{williams1992simple} as follows,
\begin{equation}
\begin{aligned}
\nabla_{\Phi} J\left(\Phi\right) 
&= E_{P\left( \textbf{w} | \Phi\right)}\left[\nabla_{\Phi} \log P\left( \textbf{w} | \Phi\right) \cdot \mathcal{R}_{val}\right]\\
\end{aligned}
\end{equation}

An empirical approximation of the above quantity is:
\begin{equation}
\begin{aligned}
\nabla_{\Phi} J\left(\Phi\right) 
&\approx  \frac{1}{|\mathcal{D}|} \sum_{(u,v) \in \mathcal{D}} \nabla_{\Phi} \log   P(w_{uv} | \Phi) \cdot \mathcal{R}_{val} \\
\end{aligned}
\end{equation}


The above update is an unbiased estimate for our gradient, but has a very high variance. In order to reduce the variance of this estimate, following \cite{zoph2016neural,pham2018efficient}, we employ a baseline function $b$, which is the moving average of the previous reward signals:

\begin{equation}\label{eq: controller_gradient}
\begin{aligned}
\frac{1}{|\mathcal{D}|}  \sum_{(u,v) \in \mathcal{D}} \nabla_{\Phi} \log   P(w_{uv} | \Phi) \cdot (\mathcal{R}_{val} - b)\\
\end{aligned}
\end{equation}




\subsection{The Controller Network}

In this subsection, we first introduce the concrete concepts for state $s$, action $a$ and reward $r$, then provide the details of model structure of the hard and soft controller network. 

\subsubsection{\textbf{State Representation}} In order to effectively and efficiently represent state, the state representation $s$ should include both arrived training data and the status of current recommendation model \cite{pmlr-v70-graves17a}. Here, we adopt three commonly-used categories of features \cite{pmlr-v70-graves17a,kumar2019reinforcement,wang2021survey}:
1) Data features, containing information for data instance, such as user and item embeddings or feature vectors; 2) Recommender model features, including the signals reflecting how well current neural network is trained. We collect one simple feature, which is the number of epochs; 3) Features that represent the combination of both data and learner model. By using these features, we aim to represent how important the arrived training data is for the current leaner. We mainly use the loss of a certain user-item pair $\ell_{uv}$, which also appears frequently in curriculum learning \cite{kumar2010self}. 

\subsubsection{\textbf{Action}} The controller has two types of actions---\textit{hard} and \textit{soft}. First, \textit{hard actions} are denoted via $a=\left\{a_{j}\right\}_{j=1}^{|\mathcal{D}|} \in\{0,1,2\}^{|\mathcal{D}|}$, representing to delete ($a_j=0$) or keep ($a_j=1$) or augment ($a_j=2$) the $j$-th instance in $\mathcal{D}$. Second, \textit{soft actions} are denoted as $a=\left\{a_{j}\right\}_{j=1}^{|\mathcal{D}|} \in {\mathcal{R}^{+}}^{|\mathcal{D}|}$, representing the training weight of the $j$-th instance in $\mathcal{D}$ as a real-value positive number. We will introduce two different network architectures to realize each of them in Section \ref{sec:controller_structure}.

\subsubsection{\textbf{Reward Function}}
We use the Area under the ROC Curve (AUC) to evaluate the performance, which is a commonly used reward in many AutoML works \cite{10.1145/3477495.3531941,zheng2022automl} (one can also select other alternative metrics, such as hit rate, NDCG, and so on), and obtain the reward on validation set since they are prediction-sensitive, i.e., the metric values will be different with a very small fluctuation on predictions \cite{allen1971mean, calders2007efficient}, so that non-trivial reward can be calculated for better update on the controller. 

\subsubsection{\textbf{Network Structure}} \label{sec:controller_structure}
Given the state representation of an input user-item pair $(u,v)$, the controller will pass it through a multi-layer perceptron and output a hidden vector $h^T_{u v}$, which is as shown below,
\begin{equation}
\begin{aligned}
    h^T_{u v} 
    =\boldsymbol{W}_{T} a_{T}\left(\ldots\left(\boldsymbol{W}_{1} a_{1}\left(concat\left(\textbf{e}_{u}, \textbf{e}_{v}, \ell_{uv}, e\right)\right)+\boldsymbol{b}_{1}\right)\ldots\right)+\boldsymbol{b}_{T}
\end{aligned}
\end{equation}
where $\boldsymbol{W}_{i}, a_{i}(\cdot), \boldsymbol{b}_{i}$ represent weights, activation functions and bias values; $e$ is the current training epoch number, $concat(\cdot)$ is the concatenation function and $\ell_{uv}=\ell \left(f( \textbf{e}_{u}, \textbf{e}_{v} \mid \Theta ), y_{uv}\right)$ for short.

In order to generate discrete actions in the hard action space, we use the multinoulli distribution (also called a categorical distribution), which is,
\begin{equation}\label{eq:AutoDenoise-H}
    w_{u v}^{hard} \sim \mathcal{M}(\textbf{softmax}(h^T_{u v}))
\end{equation}
where $\mathcal{M}$ represents the multinoulli distribution, which samples integers from $\{0, \ldots, A-1\}$ where $A$ is the number of actions given their probabilities, and $\textbf{softmax}(h^T_{u v})$ outputs the probabilities of each action. In the experiments, we define three kinds of actions to represent \{0,1,2\}, which represent "delete sample", "keep sample" and "augment sample", respectively.

For continuous actions in the soft action space, we use the Gaussian distribution. Concretely, we pass $h^T_{u v}$ through two separate output layers, one for mean and another for variance (shown in Eq. \ref{eq:mean} and Eq. \ref{eq:variance}), and get $\mu_{uv}$ and $\sigma_{u v}^2$, respectively.
\begin{equation}\label{eq:mean}
    \mu_{u v}=\boldsymbol{W}_{T+1}h^T_{u v}+\boldsymbol{b}_{T+1}
\end{equation}
\begin{equation}\label{eq:variance}
\sigma_{u v}^2=\textbf{softplus} (\boldsymbol{W}_{T+1}^{'}h^T_{u v}+\boldsymbol{b}_{T+1}^{'})
\end{equation}

Then, we use the mean and the variance to sample a continuous weight $w_{uv}$ based on Gaussian distribution, as
\begin{equation} \label{eq:AutoDenoise-S}
    w_{u v}^{soft} \sim \mathcal{N}(\mu_{u v},\sigma_{u v}^{2})
\end{equation}

Considering the current weight can be either positive or negative, which is not stable for the model training, we finally pass $w_{uv}$ through a $\textbf{softplus}$ layer to make sure it is always positive.
\begin{equation}
    \hat{w}_{u v}^{soft} = \textbf{softplus}(w_{u v})
\end{equation}

\begin{algorithm}[t]
    \textbf{Input:}  User-item interaction history $\mathcal{D}$, batch size B, learning rate $lr$ for recommender and learning rate $lr_c$ for controller. \\
    \textbf{Output:} parameters $\Theta^*$ and $\Phi^*$ \\
    Randomly initialize $\Theta$ and $\Phi$; \\
    b = 0;\\
    \While {not converged} {
        
        \For{$iteration\ =\ 1\ ...\ |\mathcal{D}|/B$}{
            Sample a minibatch with size B from $\mathcal{D}$;\\
            Calculate the weighted training loss based on Eq. \eqref{eq: weighted_loss};\\
            Fix $\Phi$, update $\Theta$ through backpropagation.
        }
        Fix $\Theta$ and pass the validation set through $f$;\\
        Get the AUC value on the validation set as reward;\\
        Update $\Phi$ based on Eq. \eqref{eq: controller_gradient};\\
        Update $b$.\\
    }
    \caption{Parameters Training for AutoDenoise}
    \label{alg:AutoDenoise}
\end{algorithm}

\subsection{Training Procedure}
We also present the detailed training procedure of the proposed framework in Algorithm \ref{alg:AutoDenoise}. 
In each epoch, there are two phases --- recommender updating phase (line 6-10) and controller updating phase (line 11-14). The former phase is similar to the normal updating except for a weighted loss. In the latter phase, we receive the reward based on the recommendation performance on the validation set, and then combine other values stored in the recommender updating phase, such as the log probabilities of the actions, to update the controller network.

\begin{table}[h]
\caption{Basic statistics of the experimental datasets. We use \#users represents the number of users, \#items the number of items and \#act. represents the total number of interactions.}
\centering
\setlength{\tabcolsep}{5pt}
\begin{adjustbox}{max width=\linewidth}
\begin{tabular}
    {lcccccc} \toprule
    Dataset & \#users & \#items & \#act. & density \\\midrule
    Movielens-100K-latest & 610 & 2,270  & 81,109 & 5.857\%\\
    Movielens-1M & 6,023 & 3,044  & 956,851 & 5.218\%\\
    Electronics & 47,726 & 30,115 & 1,143,343 & 0.079\%\\
    \bottomrule
\end{tabular}
\end{adjustbox}
\vspace{5pt}
\label{tab:dataset}
\vspace{-10pt}
\end{table}

\section{Experiments}
In this section, we first introduce the datasets, the base recommenders, the comparable baselines, then discuss and analyse the experimental results.

\subsection{Dataset Description}
To evaluate the models under different data scales, data sparsity and application scenarios, we perform experiments on three real-world public datasets.

\begin{itemize}[leftmargin=*]
    \item \textbf{Movielens dataset \cite{10.1145/2827872}:} One of the most frequently used benchmark dataset for personalized recommendation. We use \textit{Movielens-100K-latest}\footnote{\url{https://grouplens.org/datasets/movielens/latest/}} which includes about one hundred thousand user transactions (user id, item id, rating, timestamp, etc.) and was generated on September 26, 2018.
    Moreover, we choose a larger one---\textit{Movielens-1M}\footnote{https://grouplens.org/datasets/movielens/1m/} including over one million user-item interactions.
    
    \item \textbf{Amazon dataset \cite{he2016ups}:} 
    This dataset contains user reviews on products in Amazon e-commerce system.\footnote{\url{https://nijianmo.github.io/amazon/index.html}}
    It has 29 sub-datasets corresponding to 29 product categories. 
    We adopt \textit{Electronics} dataset to evaluate our method.
\end{itemize}

The original data is huge and highly sparse, especially Amazon dataset, which makes it challenging to evaluate.
Therefore, similar to previous work \cite{he2017neural, kang2018self}, we filter out users and items with fewer than ten interactions. Then, for each dataset, we sort the records of each user based on the timestamp, and split the records into training, validation, and testing sets chronologically by 4:1:1. 

Moreover, to mimic the real-world scenarios as close as possible (large-scale implicit feedback and only a small volume of explicit feedback), we make some modifications to the traditional training, validation, testing process.
First, for training set, we keep all observed interactions in the training set and keep them implicit (i.e., without using any explicit rating information).
Second, for validation- and testing sets, we only keep users' true-positive interactions since we need to use a small number of noise-free validation set to guide the model training, and evaluate the recommendation performance on a holdout clean testing set with only true-positive interactions kept, i.e., the evaluation focuses on recommending more satisfying items to users.
Following \cite{wang2021denoising}, we use the ground-truth rating scores to filter out false-positive data samples in the original validation- and testing sets.
More specific, an interaction is identified as false-positive or true-positive according to the explicit feedback, and we define an interaction as false-positive if its rating score ([1, 5]) < 3.

Some basic statistics of the filtered experimental datasets are shown in Table \ref{tab:dataset}, including number of users, number of items, number of interactions and density.

\begin{table*}[]
\footnotesize
\caption{ Summary of the performance on three benchmark datasets. 
We evaluate $Precision$, $Recall$, $F_1$ and $NDCG$, in percentage (\%) values (\% symbol is omitted in the table for clarity), whiles $K$ is the length of recommendation list. 
When AutoDenoise-augmented methods are the best, i.e., AutoDenoise-H  or AutoDenoise-S, its improvements against the best baseline are significant at p < 0.01. 
Bold scores are used for the largest values.
}
\centering
\begin{adjustbox}{max width=\linewidth}
\setlength{\tabcolsep}{7pt}
\begin{tabular}
    {m{2.66cm} ccc ccc ccc ccc} \toprule
    \multirow{2.5}{*}{Methods} 
    & \multicolumn{3}{c}{Precision (\%) $\uparrow$}
    & \multicolumn{3}{c}{Recall (\%) $\uparrow$} 
    & \multicolumn{3}{c}{F1 (\%) $\uparrow$} 
    & \multicolumn{3}{c}{NDCG (\%) $\uparrow$} 
    \\\cmidrule(lr){2-4} \cmidrule(lr){5-7} \cmidrule(lr){8-10} \cmidrule(lr){11-13} 
 & K=10 & K=20 & K=50 & K=10 & K=20 & K=50 & K=10 & K=20 & K=50 & K=10 & K=20 & K=50 \\\midrule 
 \multicolumn{13}{c}{Movielens-100K-latest} \\\midrule
MF-Default & 0.9688 & 0.7881 & 0.6108 & 9.6880 & 15.763 & 30.541 & 0.0176 & 0.0150 & 0.0119 & 4.4379 & 6.0021 & 8.8955 \\
MF-Heuristic & 0.9523 & 0.7471 & 0.5911 & 9.5238 & 14.942 & 29.556 & 0.0173 & 0.0142 & 0.0115 & 4.9292 & 6.2640 & 9.1015 \\
MF-ADT-TL & 1.0016 & 0.8949 & 0.7060 & 10.016 & 17.898 & 35.303 & 0.0182 & 0.0170 & 0.0138 & 5.0444 & 7.0297 & 10.430 \\
MF-ADT-RL & \textbf{1.0673} & 0.8292 & 0.7027 & \textbf{10.673} & 16.584 & 35.139 & \textbf{0.0194} & 0.0157 & 0.0137 & \textbf{5.3187} & 6.7970 & 10.436 \\
MF-AutoDenoise-H & 0.9523 & 0.8292 & \textbf{0.7224} & 9.5238 & 16.584 & \textbf{36.124} & 0.0173 & 0.0157 & \textbf{0.0141} & 4.8617 & 6.6550 & \textbf{10.515} \\
MF-AutoDenoise-S & 0.9688 & \textbf{0.9113} & 0.6765 & 9.6880 & \textbf{18.226} & 33.825 & 0.0176 & \textbf{0.017}3 & 0.0132 & 5.1248 & \textbf{7.2770} & 10.314 \\
\cmidrule(lr){2-13}

GRU4Rec-Default & 0.8538 & 0.7881 & 0.5451 & 8.5385 & 15.763 & 27.257 & 0.0155 & 0.0150 & 0.0106 & 4.2018 & 6.0134 & 8.2861 \\
GRU4Rec-Heuristic & 0.8702 & 0.6978 & 0.5845 & 8.7027 & 13.957 & 29.228 & 0.0158 & 0.0132 & 0.0114 & 4.2326 & 5.5695 & 8.5895 \\
GRU4Rec-ADT-TL & 0.8374 & 0.6814 & 0.5977 & 8.3743 & 13.628 & 29.885 & 0.0152 & 0.0129 & 0.0117 & 4.2777 & 5.5915 & 8.8293 \\
GRU4Rec-ADT-RL & 0.6075 & 0.5911 & 0.5057 & 6.0755 & 11.822 & 25.287 & 0.0110 & 0.0112 & 0.0099 & 2.4933 & 3.9205 & 6.5986 \\
GRU4Rec-AutoDenoise-H & 0.7881 & 0.7635 & 0.6338 & 7.8817 & 15.270 & 31.691 & 0.0143 & 0.0145 & 0.0124 & 4.3511 & 6.2202 & 9.4386 \\
GRU4Rec-AutoDenoise-S & \textbf{0.9195} & \textbf{0.8292} & \textbf{0.6469} & \textbf{9.1954} & \textbf{16.584} & \textbf{32.348} & \textbf{0.0167} & \textbf{0.0157} & \textbf{0.0126} & \textbf{4.8351} & \textbf{6.6880} & \textbf{9.7809} \\
\cmidrule(lr){2-13}

SASRec-Default & 0.9852 & 0.7389 & 0.5221 & 9.8522 & 14.778 & 26.108 & 0.0179 & 0.0140 & 0.0102 & 6.3683 & 7.5960 & 9.8342 \\
SASRec-Heuristic & 1.1494 & 0.8456 & 0.5188 & 11.494 & 16.912 & 25.944 & 0.0208 & 0.0161 & 0.0101 & 7.6434 & 8.9963 & 10.824 \\
SASRec-ADT-TL & 0.8045 & 0.5008 & 0.2922 & 8.0459 & 10.016 & 14.614 & 0.0146 & 0.0095 & 0.0057 & 7.1405 & 7.6332 & 8.5357 \\
SASRec-ADT-RL & 1.1986 & 0.8538 & 0.5353 & 11.986 & 17.077 & 26.765 & 0.0217 & 0.0162 & 0.0104 & 8.1410 & 9.4106 & 11.321 \\
SASRec-AutoDenoise-H & 1.1822 & 0.7635 & 0.5254 & 11.822 & 15.270 & 26.272 & 0.0214 & 0.0145 & 0.0103 & 7.8775 & 8.7565 & 10.942 \\
SASRec-AutoDenoise-S & \textbf{1.4285} & \textbf{1.0098} & \textbf{0.6765} & \textbf{14.285} & \textbf{20.197} & \textbf{33.825} & \textbf{0.0259} & \textbf{0.0192} & \textbf{0.0132} & \textbf{9.7465} & \textbf{11.248} & \textbf{13.967} \\
\midrule

\multicolumn{13}{c}{Movielens-1M} \\\midrule
MF-Default & 0.8535 & 0.7198 & 0.5626 & 8.5353 & 14.397 & 28.130 & 0.0155 & 0.0137 & 0.0110 & 4.1121 & 5.5846 & 8.2822 \\
MF-Heuristic & 0.7871 & 0.6766 & 0.5363 & 7.8711 & 13.533 & 26.818 & 0.0143 & 0.0128 & 0.0105 & 3.8129 & 5.2273 & 7.8491 \\
MF-ADT-TL & 0.7820 & 0.6984 & 0.5183 & 7.8209 & 13.969 & 25.918 & 0.0142 & 0.0133 & 0.0101 & 3.9322 & 5.4758 & 7.8220 \\
MF-ADT-RL & 0.8291 & 0.7037 & 0.5445 & 8.2912 & 14.074 & 27.225 & 0.0150 & 0.0134 & 0.0106 & 4.1125 & 5.5588 & 8.1520 \\
MF-AutoDenoise-H & 0.8684 & 0.7480 & 0.5569 & 8.6848 & 14.961 & 27.847 & 0.0157 & 0.0142 & 0.0109 & 4.3586 & 5.9325 & 8.4701 \\
MF-AutoDenoise-S & \textbf{0.9182} & \textbf{0.8145} & \textbf{0.6130} & \textbf{9.1829} & \textbf{16.290} & \textbf{30.654} & \textbf{0.0166} & \textbf{0.0155} & \textbf{0.0120} & \textbf{4.6667} & \textbf{6.4542} & \textbf{9.2802} \\
\cmidrule(lr){2-13}

GRU4Rec-Default & 2.0209 & 1.5318 & 0.9614 & 20.209 & 30.637 & 48.073 & 0.0367 & 0.0291 & 0.0188 & 10.318 & 12.946 & 16.393 \\
GRU4Rec-Heuristic & 1.4314 & 1.1366 & 0.7847 & 14.314 & 22.733 & 39.239 & 0.0260 & 0.0216 & 0.0153 & 7.0436 & 9.1565 & 12.422 \\
GRU4Rec-ADT-TL & 2.1912 & 1.5981 & 0.9858 & 21.912 & 31.963 & 49.294 & 0.0398 & 0.0304 & 0.0193 & 11.287 & 13.806 & 17.250 \\
GRU4Rec-ADT-RL & 2.1233 & 1.6295 & 0.9970 & 21.233 & 32.590 & 49.851 & 0.0386 & 0.0310 & 0.0195 & 11.020 & 13.876 & 17.302 \\
GRU4Rec-AutoDenoise-H & 2.3862 & 1.7577 & 1.0318 & 23.862 & 35.154 & 51.594 & 0.0433 & 0.0334 & 0.0202 & 12.285 & 15.125 & 18.386 \\
GRU4Rec-AutoDenoise-S & \textbf{2.4211} & \textbf{1.7701} & \textbf{1.0577} & \textbf{24.211} & \textbf{35.403} & \textbf{52.889} & \textbf{0.0440} & \textbf{0.0337} & \textbf{0.0207} & \textbf{13.078} & \textbf{15.899} & \textbf{19.377} \\
\cmidrule(lr){2-13}

SASRec-Default & 1.6124 & 1.1383 & 0.7160 & 16.124 & 22.766 & 35.802 & 0.0293 & 0.0216 & 0.0140 & 10.256 & 11.931 & 14.493 \\
SASRec-Heuristic & 1.6921 & 1.1665 & 0.7157 & 16.921 & 23.331 & 35.785 & 0.0307 & 0.0222 & 0.0140 & 10.496 & 12.110 & 14.556 \\
SASRec-ADT-TL & 1.6199 & 1.1757 & 0.7249 & 16.199 & 23.515 & 36.248 & 0.0294 & 0.0223 & 0.0142 & 10.043 & 11.880 & 14.390 \\
SASRec-ADT-RL & 1.7070 & 1.1478 & 0.7197 & 17.070 & 22.957 & 35.986 & 0.0310 & 0.0218 & 0.0141 & 11.042 & 12.529 & 15.106 \\
SASRec-AutoDenoise-H & 1.6140 & 1.1017 & 0.7014 & 16.140 & 22.035 & 35.071 & 0.0293 & 0.0209 & 0.0137 & 10.628 & 12.111 & 14.677 \\
SASRec-AutoDenoise-S & \textbf{1.8930} & \textbf{1.2811} & \textbf{0.7818} & \textbf{18.930} & \textbf{25.622} & \textbf{39.090} & \textbf{0.0344} & \textbf{0.0244} & \textbf{0.0153} & \textbf{12.065} & \textbf{13.739} & \textbf{16.397} \\
\midrule

\multicolumn{13}{c}{Electronics} \\\midrule
MF-Default  & 0.0800 & 0.0712 & 0.0581 & 0.8004 & 1.4248 & 2.9083 & 0.0014 & 0.0013 & 0.0011 & 0.4171 & 0.5738 & 0.8652 \\
MF-Heuristic & 0.0865 & 0.0748 & \textbf{0.0643} & 0.8653 & 1.4960 & \textbf{3.2163} & 0.0015 & 0.0014 & 0.0012 & 0.4449 & 0.6023 & \textbf{0.9396} \\
MF-ADT-TL & 0.0854 & 0.0755 & 0.0611 & 0.8548 & 1.5107 & 3.0570 & 0.0015 & 0.0014 & 0.0011 & 0.4207 & 0.5836 & 0.8870 \\
MF-ADT-RL & 0.0735 & 0.0692 & 0.0561 & 0.7354 & 1.3850 & 2.8098 & 0.0013 & 0.0013 & 0.0011 & 0.3654 & 0.5277 & 0.8067 \\
MF-AutoDenoise-H & 0.0865 & 0.0733 & 0.0615 & 0.8653 & 1.4667 & 3.0759 & 0.0015 & 0.0013 & 0.0012 & 0.4406 & 0.5908 & 0.9076 \\
MF-AutoDenoise-S & \textbf{0.0875} & \textbf{0.0762} & 0.0625 & \textbf{0.8758} & \textbf{1.5254} & 3.1283 & \textbf{0.0015} & \textbf{0.0014} & \textbf{0.0012} & \textbf{0.4564} & \textbf{0.6185} & 0.9335 \\
\cmidrule(lr){2-13}

GRU4Rec-Default & 0.2095 & 0.1697 & 0.1220 & 2.0953 & 3.3944 & 6.1016 & 0.0038 & 0.0032 & 0.0023 & 1.0350 & 1.3612 & 1.8950 \\
GRU4Rec-Heuristic & 0.2338 & 0.1835 & 0.1260 & 2.3383 & 3.6710 & 6.3027 & 0.0042 & 0.0034 & 0.0024 & 1.2862 & 1.6214 & 2.1401 \\
GRU4Rec-ADT-TL & 0.2610 & 0.1947 & 0.1351 & 2.6107 & 3.8952 & 6.7574 & 0.0047 & 0.0037 & 0.0026 & 1.4820 & 1.8037 & 2.3671 \\
GRU4Rec-ADT-RL & 0.2409 & 0.1884 & 0.1351 & 2.4096 & 3.7695 & 6.7553 & 0.0043 & 0.0035 & 0.0026 & 1.2427 & 1.5843 & 2.1712 \\
GRU4Rec-AutoDenoise-H &  0.2772 &  0.2073 &  0.1388 &  2.7721 &  4.1466 &  6.9418 &  0.0050 &  0.0039 & 0.0027 & 1.4782 &  1.8221 &  2.3738 \\
GRU4Rec-AutoDenoise-S & \textbf{0.3073} & \textbf{0.2280} & \textbf{0.1511} & \textbf{3.0738} & \textbf{4.5615} & \textbf{7.5578} & \textbf{0.0055} & \textbf{0.0043} & \textbf{0.0029} & \textbf{1.8079} & \textbf{2.1827} & \textbf{2.7724} \\
\cmidrule(lr){2-13}

SASRec-Default & 0.4854 & 0.2964 & 0.1662 & 4.8548 & 5.9298 & 8.3122 & 0.0088 & 0.0056 & 0.0032 & 3.8209 & 4.0909 & 4.5591 \\
SASRec-Heuristic & 0.4521 & 0.2380 & 0.1039 & 4.5217 & 4.7606 & 5.1985 & 0.0082 & 0.0045 & 0.0020 & 4.1210 & 4.1812 & 4.2674 \\
SASRec-ADT-TL & 0.4695 & 0.3022 & \textbf{0.1734} & 4.6956 & 6.0450 & \textbf{8.6726} & 0.0085 & 0.0057 & 0.0034 & 3.2154 & 3.5550 & 4.0740 \\
SASRec-ADT-RL & 0.5068 & 0.3040 & 0.1654 & 5.0686 & 6.0806 & 8.2744 & 0.0092 & 0.0057 & 0.0032 & 3.9454 & 4.2000 & 4.6300 \\
SASRec-AutoDenoise-H & 0.4817 & 0.2854 & 0.1525 & 4.8171 & 5.7097 & 7.6291 & 0.0087 & 0.0054 & 0.0029 & 3.9232 & 4.1473 & 4.5256 \\
SASRec-AutoDenoise-S & \textbf{0.5227} & \textbf{0.3052} & 0.1583 & \textbf{5.2278} & \textbf{6.1058} & 7.9161 & \textbf{0.0095} & \textbf{0.0058} & \textbf{0.0031} & \textbf{4.2245} & \textbf{4.4449} & \textbf{4.8023} \\
\bottomrule
\end{tabular}\label{tab:result_ml}
\end{adjustbox}
\end{table*}

\subsection{Experimental Setup}\label{sec:experimental_setup}
\subsubsection{\textbf{Base Recommenders.}}
We compare our proposed method with the following baselines, including both traditional and reinforcement learning  based recommendation models. 

\begin{itemize}
\item {\bf MF}: Collaborative Filtering based on matrix factorization (MF) \cite{koren2009matrix} is a representative method for rating prediction task. 
In our implementation, we turn the rating prediction task into ranking prediction and use Softmax loss.


\item {\bf GRU4Rec \cite{hidasi2015session}}: It applies Gated Recurrent Units (GRU) to model user click sequence for session-based recommendation. It models user interactions with items, such as clicks, by analyzing sequences of interactions. In our implementation, we represent the items using embedding vectors rather than one-hot vectors.

\item {\bf SASRec \cite{kang2018self}}: It is a deep learning model for session-based recommendation. It uses self-attention mechanism to weight the importance of different items in a user's interaction history, allowing it to better capture the user's preferences. 

\end{itemize}

\begin{figure*}[h]
\mbox{
\centering
    \subfigure[Results in Movielens-100K-latest]{\label{fig:learned_weights-ml-100k}
        \includegraphics[width=0.33\textwidth]{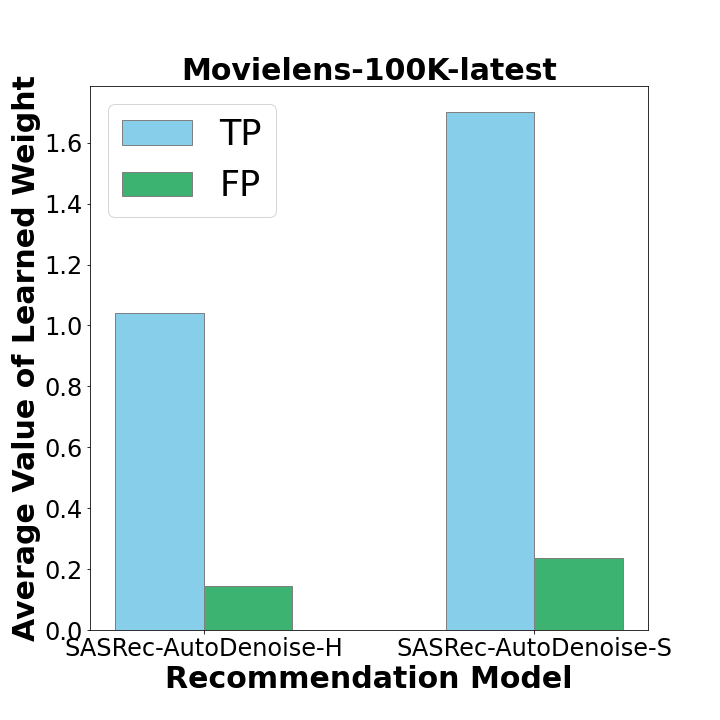}}
    \hspace{-5pt}
    \subfigure[Results in Movielens-1M]{\label{fig:learned_weights-ml-1m}
        \includegraphics[width=0.33\textwidth]{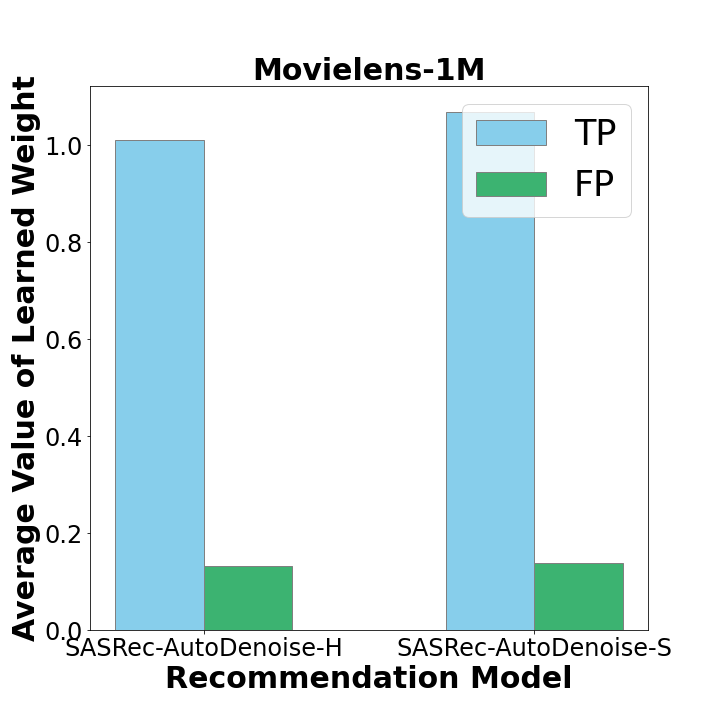}}
    \hspace{-5pt}
    \subfigure[Results in Electronics]{\label{fig:learned_weights-electronics}
        \includegraphics[width=0.33\textwidth]{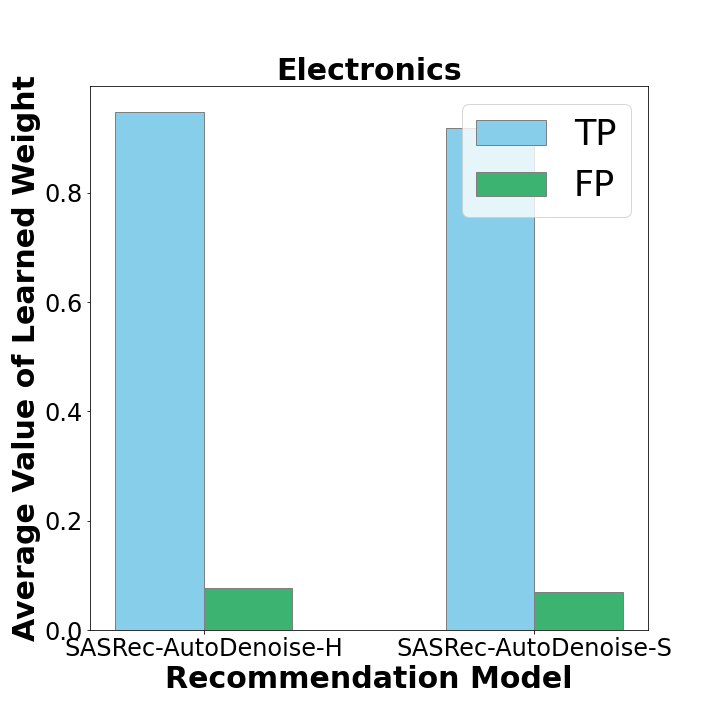}}
    
}
\caption{The average values of learned weights for ground-truth true-positive and false-positive training samples in all three datasets. The legend indicates that TP (blue bar) stands for true-positive and FP (green bar) stands for false-positive data samples. The x-axis includes AutoDenoise-H and AutoDenoise-S with SASRec as the base ranker. The y-axis represents the average values of the learned weights.}
\label{fig:learned_weights}
\end{figure*}

\subsubsection{\textbf{Reprocessing Schemes.}}
To evaluate the effectiveness of the reweighting scheme, we incorporate the following four reweighting schemes with all base recommenders.

\begin{itemize}
    \item {\bf Default}: The default scheme involves no sample reweighting.
    
    \item {\bf Heuristic}:  The heuristic scheme involves positive sample upweighting based on a pre-defined heuristic function. Specifically, following \cite{ren2018learning,hu2019learning}, we evaluate the commonly-used proportion method that weights each example by the inverse frequency, where frequency is the number of times an item appears in the training data.
    

    \item {\bf Adaptive Denoising Training with Truncated Loss (ADT-TL)}: \citet{wang2021denoising} proposed ADT strategies for recommender systems, which dynamically prunes the large-loss interactions along the training process. Specifically, there are two paradigms formulating the training loss, and ART-TL using Truncated Loss, which truncates the loss values of hard interactions to 0 with a dynamic threshold function.

    \item {\bf Adaptive Denoising Training with Reweighted Loss (ADT-RL)}: This method is another ADT strategy proposed by \citet{wang2021denoising}.
    Specifically, ADT-RL using Reweighted Loss, which assigns hard interactions with smaller weights based on a variant of Focal Loss \cite{lin2017focal}. 
    
    \item {\bf Hard AutoDenoise (AutoDenoise-H)}: AutoDenoise-H is our proposed AutoDenoise scheme with discrete action space.
    
    \item {\bf Soft AutoDenoise (AutoDenoise-S)}: AutoDenoise-S is our proposed AutoDenoise scheme with continuous action space.
\end{itemize}

\subsubsection{\textbf{Implementation Details}}
We implement MF-Default, MF-Heuristic, MF-ADT-TL, MF-ADT-RL, GRU4Rec-Default, GRU4Rec-Heuristic, GRU4Rec-ADT-TL, GRU4Rec-ADT-RL, SASRec-Default, SASRec-Heuristic, SASRec-ADT-TL, SASRec-ADT-RL using \textit{Pytorch} with Adam optimizer.
For all of them, we consider latent dimensions $d$ from \{32, 64, 128\}, 
learning rate $lr$ from \{1e-2, 5e-3, 1e-3, 5e-4, 1e-4, 5e-5\}, L2 penalty is chosen from \{1e-3, 1e-4, 1e-5, 1e-6\}, batch size is select from \{256, 512, 1024, 2048\}, and use Softmax loss. 
For GRU4Rec and SASRec, since they are sequence models, we use a sliding window over the items with maximum 10 items.
We tune the hyper-parameters using the validation set and terminate training when the performance on the validation set does not change within 6 epochs.

Moreover, for AutoDenoise-augmented recommender systems, e.g., MF-AutoDenoise-H, MF-AutoDenoise-S, GRU4Rec-AutoDenoise-H, GRU4Rec-AutoDenoise-S, SASRec-AutoDenoise-H and SASRec-AutoDenoise-S, we design the controller network by using a Multi-Layer Perceptron (MLP) with two layers and ReLU function as the embedding layer, and the output layer of AutoDenoise-H follows Eq. \ref{eq:AutoDenoise-H}, and that of AutoDenoise-S follows Eq. \ref{eq:AutoDenoise-S}. We optimize the controller using Adam optimizer. Besides, we consider the learning rate $lr$ from \{1e-2, 5e-3, 1e-3, 5e-4, 1e-4, 5e-5\}, and the L2 penalty is chosen from \{1e-3, 1e-4, 1e-5, 1e-6\}.

\subsubsection{\textbf{Evaluation Metrics.}}
We select several most commonly used top-$K$ ranking metrics to evaluate each model's recommendation performance, including \textbf{Precision}, \textbf{Recall}, \textbf{F1 Score}, and \textbf{NDCG}, with different lengths of recommendation lists, i.e., @10, @20, @50.

\begin{figure*}[t]
\mbox{
\hspace{-5pt}
\centering
    \subfigure[Results in Movielens-100K-latest]{\label{fig:ablation_study_ml-100k}
        \includegraphics[width=0.34\textwidth]{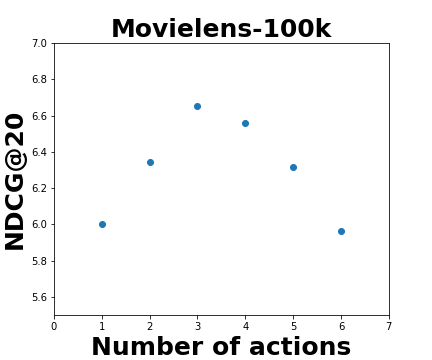}}
    \hspace{-5pt}
    \subfigure[Results in Movielens-1M]{\label{fig:ablation_study_ml-1m}
        \includegraphics[width=0.34\textwidth]{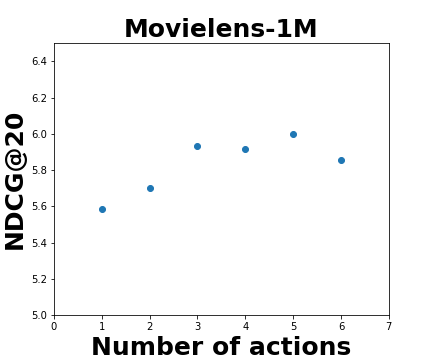}}
    \hspace{-5pt}
    \subfigure[Results in Electronics]{\label{fig:ablation_study_electronics}
        \includegraphics[width=0.34\textwidth]{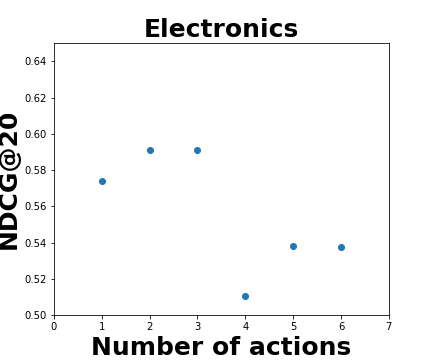}}
    
}
\caption{Relationship between number of actions and recommendation performance for MF-AutoDenoise-H on three datasets. $x$-axis is the number of actions chosen for hard AutoDenoise method; $y$-axis represents NDCG@20 on test set after convergence.}
\label{fig:ablation}
\vspace{-6pt}
\end{figure*}

\subsection{Experimental Results}
The major experimental results on Movielens-100K-lateset, Movielens-1M and Electronics datasets are shown in Table \ref{tab:result_ml}.
We analyze and discuss the results in terms of the following perspectives.

\subsubsection{\bf Shallow models vs. Deep models}
Among all the base recommenders, we can see that both deep models (GRU4Rec-Default, SASRec-Default) are better than the shallow model (MF-Default) in most cases. Specifically, when averaging across all metrics on all three datasets, GRU4Rec-Default gets 45.86\% improvement than MF-Default, and SASRec-Default even achieves 136.77\% improvement.
The greatest improvement is achieved by SASRec-Default on Electronics dataset on NDCG@10, which is 816.0\%. 
These observations verify the effectiveness of deep recommendation models and their abilities to learn more accurate user preferences.
The reason behind is that MF uses a user’s historical interaction to learn the static preference, while sequential recommendation models leverage the fact that the next behavior of a user not only depends on the static long-term preference, but also relies on the current intent \cite{hidasi2015session,kang2018self}. 
Another explanation is related to data density. 
As we can see that for deep sequential recommenders, the sparser the dataset is, the greater the improvement is. 

\subsubsection{\bf Default scheme vs. Reweighting schemes}
Among all the reweighting schemes, we can see that reweighting methods are able to improve the default recommendation performance in many cases. 
For instance, when averaging across all metrics and all base rankers, Heuristic gets 3.24\% on Movielens-100K-latest and 0.74\% improvement on Electronics, ADT-TL gets 0.26\% on Movielens-1M and 7.78\% on Electronics, and ADT-RL, the strongest baseline, gets 1.32\% on Movielens-100K-latest, 2.12\% on Movielens-1M and 3.44\% on Electronics. Moreover, our proposed AutoDenoise-augmented methods are always better than the default methods. Specifically, when averaging across all metrics on all three datasets, AutoDenoise-H gets overall 7.29\% improvement than Default, with 7.32\% on Movielens-100K-latest, 5.20\% on Movielens-1M and 9.35\% on Electronics; and AutoDenoise-S gets overall 17.44\% improvement, with 20.75\% on Movielens-100K-latest, 13.83\% on Movielens-1M and 17.73\% on Electronics. 

\subsubsection{\bf Hard policy vs. Soft policy}
By comparing both ADT \cite{wang2021denoising} and AutoDenoise's discrete and continuous versions in Table \ref{tab:result_ml}, we can easily find that AutoDenoise-S outperforms AutoDenoise-H, and ADT-RL outperforms ADT-TL in most cases. For example, ADT-RL gets 11.37\% improvement than ADT-RL on Movielens-100K-latest, 1.98\% on Movielens-1M. Similarly, AutoDenoise-S achieves 12.22\% improvement than AutoDenoise-H on Movielens-100K-latest, 8.49\% on Movielens-1M, and 7.26\% on Electronics when averaging across all metrics, which shows the strong generalizability of continuous action space. Such action space is able to provide more differences of interaction importance so as to better denoise the data and improve the recommendation performance. Additionally, when averaging across all metrics on three datasets using different base rankers, AutoDenoise-S achieves 4.47\% improvement than AutoDenoise-H on MF, 7.80\% on GRU4Rec, and 16.18\% on SASRec, which also shows a stronger generalizability to fit different base rankers.

\subsection{In-depth Analysis}
To gain deeper understandings of the inner mechanism of the proposed AutoDenoise framework, we further analyze the learned weights after model convergence and examine the selection of action space for hard AutoDenoise.

\subsubsection{\bf Study of Learned Weights}
In order to study the cause of the improvements observed in AutoDenoise, we collect the weights assigned to each training sample after model convergence, and analyse them by comparing with the ground-truth labels.
Specifically, we first label each training sample to either true-positive or false-positive based on its rating score. 
Then, we calculate the averaged weights for both true-positive sample group and false-positive sample group, and plot the results from all three datasets in Fig. \ref{fig:learned_weights}.
For the sake of convenience, all the results are from SASRec-AutoDenoise-H and SASRec-AutoDenoise-S on Movielens-100K-latest, Movielens-1M and Electronics datasets. 
Similar patterns can also be found using other recommendation methods.

In Fig. \ref{fig:learned_weights}, we can easily see that the average weights learned for true-positive samples are much greater than these for false-positive ones, which indicates that the proposed method is able to mitigate the noise in the training data and achieve better recommendation performance. Specifically, the average value of learned weights of true-positive training samples is 1.114 when averaging across all datasets, and that of false-positive ones is 0.132.

\subsubsection{\bf Study of Action Space}
In order to explain some of our original design choices, we study the relationship between the size of the action space for hard AutoDenoise and its performance. For convenience, we present results from MF-AutoDenoise-H on Movielens-100K-latest, Movielens-1M and Electronics by setting different numbers of actions for the controller network. Similar patterns can also be found using other base rankers.
As shown in Fig. \ref{fig:ablation}, the recommendation performance (NDCG@20 on y-axis) increases as we increase the discrete action space, reaching a peak at 3 on Movielens-100K and Electronics, or at 5 on Movielens-1M, then decreases. On the x-axis, 1 means only keeping the sample, 2 is either deleting or keeping the sample, 3 represents \{delete the sample, keep the sample, duplicate the sample two times\} and so on. This observation explains why we choose to expand the action space for data augmenting in AutoDenoise-H or set weights greater than one in AutoDenoise-S, rather than directly following the traditional design of curriculum learning. Even if two true-positive items are both preferred by a certain user, there is still a preference difference between them, similar for the false-positive ones. Using a larger action space or even a continuous space provides more freedom of choices, thus, resulting in better performance.

\section{Conclusion}
In this work, we aim to deal with the noisy nature of implicit feedback for model training in recommendation.
Specifically, we select a small number of explicit feedback as validation set to guide the recommender training process, and propose an Automated Data Denoising framework (AutoDenoise) based on reinforcement learning to automatically and adaptively learn the most appropriate weights for each implicit data sample under the guidance of the noise-free validation set. 
We conduct thorough experiments, which indicate that the proposed framework is capable of boosting the performance of the state-of-the-art recommendation algorithms on several benchmark datasets.

\balance
\bibliographystyle{ACM-Reference-Format}
\bibliography{paper}

\end{document}